\documentclass[12pt,twocolumn,showpacs,preprintnumbers,amsmath,amssymb,superscriptaddress]{aastex63}
\usepackage{amsmath}
\usepackage{graphicx}
\usepackage{amsfonts}
\usepackage{amssymb}
\usepackage{bm}
\usepackage{color}%
\usepackage{ulem}
\usepackage{subfigure}
\usepackage{mathrsfs}
\usepackage{hyperref}

\begin{document}
	
	\title{Molecular Gas Excitation of the Massive Dusty Starburst CRLE and the Main-Sequence Galaxy HZ10 at z=5.7 in the COSMOS Field}
	
	\author{Daniel Vieira}
	
	\affiliation{I. Physikalisches Institut, Universit{\"a}t zu K{\"o}ln, Z{\"u}lpicher Stra{\ss}e 77, 50937 K{\"o}ln, Germany}
	
	\affiliation{Department of Astronomy, Cornell University, Space Sciences Building, Ithaca, NY 14853, USA}
	
	\author{Dominik A. Riechers}
	
	\affiliation{I. Physikalisches Institut, Universit{\"a}t zu K{\"o}ln, Z{\"u}lpicher Stra{\ss}e 77, 50937 K{\"o}ln, Germany}
	
	\author{Riccardo Pavesi}
	
	\affiliation{Department of Astronomy, Cornell University, Space Sciences Building, Ithaca, NY 14853, USA}
	
	\author{Andreas L. Faisst}
	
	\affiliation{IPAC, California Institute of Technology, 1200 East California Boulevard, Pasadena, CA 91125, USA}
	
	\author{Eva Schinnerer}
	
	\affiliation{Max Planck Institute for Astronomy, K{\"o}nigstuhl 17, 69117 Heidelberg, Germany}
	
	\author{Nicholas Z. Scoville}
	
	\affiliation{Astronomy Department, California Institute of Technology, MC 249-17, 1200 East California Boulevard, Pasadena, CA 91125, USA}
	
	\author{Gordon J. Stacey}
	
	\affiliation{Department of Astronomy, Cornell University, Space Sciences Building, Ithaca, NY 14853, USA}
	
	\begin{abstract}
		We report CO(5$\rightarrow$4) and CO(6$\rightarrow$5) line observations in the dusty starbursting galaxy CRLE ($z = 5.667$) and the main-sequence (MS) galaxy HZ10 ($z = 5.654$) with the Northern Extended Millimeter Array (NOEMA). CRLE is the most luminous $z>5$ starburst in the COSMOS field and HZ10 is the most gas-rich ``normal'' galaxy currently known at $z>5$. We find line luminosities for CO(5$\rightarrow$4) and CO(6$\rightarrow$5) of (4.9 $\pm$ 0.5) and (3.8 $\pm$ 0.4) $\times$ 10$^{10}$ K km s$^{-1}$ pc$^{2}$ for CRLE and upper limits of $< 0.76$ and $< 0.60$ $\times$ 10$^{10}$ K km s$^{-1}$ pc$^{2}$ for HZ10, respectively. The CO excitation of CRLE appears comparable to other $z>5$ dusty star-forming galaxies (DSFGs). For HZ10, these line luminosity limits provide the first significant constraints of this kind for a MS galaxy at $z > 5$. We find the upper limit of $L'_{5\rightarrow4}/L'_{2\rightarrow1}$ in HZ10 could be similar to the average value for MS galaxies around $z\approx 1.5$, suggesting that MS galaxies with comparable gas excitation may already have existed one billion years after the Big Bang. For CRLE we determine the most likely values for the H$_2$ density, kinetic temperature and dust temperature based on excitation modeling of the CO line ladder. We also derive a total gas mass of $(7.1 \pm 1.3) \times 10^{10} M_\odot$. Our findings provide some of the currently most detailed constraints on the gas excitation that sets the conditions for star formation in a galaxy protocluster environment at $z > 5$.
		\vspace{2cm}
	\end{abstract}
	
	\section{Introduction}
	
	In recent years a significant population of high-redshift ($z>5$) hyper-luminous galaxies has been discovered at sub-millimeter wavelengths (eg. Riechers et al. \hyperlink{R1}{2010}, \hyperlink{R2}{2013}, \hyperlink{R3}{2017}; Weiss et al. \hyperlink{We}{2013}; Strandet et al. \hyperlink{S1}{2016}, \hyperlink{S2}{2017}; see Riechers et al. \hyperlink{R4}{2020} for a recent summary). Likely driven by major mergers, this population represents a small fraction of ``sub-millimeter" galaxies (SMGs) and is typified by star formation rates $>$ 1000 $M_\odot$ yr$^{-1}$ in compact regions $<$ 10 kpc in diameter. Such extreme star-forming conditions are common in the so-called ``Hyper-Luminous Infrared Galaxies" (HyLIRGs), galaxies defined by IR luminosities exceeding 10$^{13}$ $L_\odot$. 
	
	They may play a key role in the evolution of massive elliptical galaxies in the early Universe: the discovery of large, passive ellipticals at $z > 2$ with stellar masses $>$ 2$\times 10^{11} M_\odot$ cannot be explained by ``conventional" SMGs as their mean molecular gas mass (5$\times 10^{10} M_\odot$) is inadequate (Damjanov et al. \hyperlink{D}{2009}, Fu et al. \hyperlink{F}{2013}). However, HyLIRGs have the capacity to form such massive elliptical galaxies relatively early in the Universe's history. 
	
	In this sense, observations of the most luminous starbursts at the highest redshifts may allow us to better understand the formation history of elliptical galaxies observed at lower redshifts, with sub-mm interferometry of the gas and dust playing an especially important role in uncovering how exactly this formation takes place.
	
	At all redshifts, starburst galaxies have been found to contribute approximately 10$\%$ to the total cosmic star formation rate density (Carilli \& Walter \hyperlink{CW}{2013}). Despite this, relatively few theoretical models of galaxy formation and evolution account for the contribution of the starburst population (Wang et al. \hyperlink{W}{2018}).
	
	In the study of high-redshift starbursts, molecular gas observations are of central importance as they can help identify the physical conditions present in a star-forming region. While low-$J$ CO transitions trace both cold and warm gas, higher-$J$ CO transitions preferentially trace warmer, denser regions which may be associated with higher star formation rates (Carilli \& Walter \hyperlink{CW}{2013}). Observations of higher-$J$ lines are thus of high importance in the study of extreme star-forming environments such as those present in HyLIRGs.
	
	One of the most distant HyLIRGs currently known was recently found serendipitously, the dusty star-forming galaxy ``CRLE" at $z=5.667$ (J2000 $10^\mathrm{h}00^\mathrm{m}59^\mathrm{s}.2$, $1^\mathrm{o}33'06''.6$) with a FIR luminosity\footnote{The FIR luminosity is defined by integrating between 42.5 and 122.5 $\mu$m} of $(1.6 \pm 0.1) \times 10^{13} L_\odot$ (Pavesi et al. \hyperlink{P}{2018}, hereafter P18). At the time of its discovery, CRLE was the highest-redshift starburst in the COSMOS field; it remains the intrinsically most infrared-luminous galaxy in COSMOS at $z>5$. Found to have a massive cold molecular gas reservoir by the observation of CO(2$\rightarrow$1) line emission, it has a star formation rate (SFR) exceeding 1500 $M_\odot$ yr$^{-1}$ in a region only $\sim$ 3 kpc in diameter. This results in a high SFR surface density of $\sim$ 100 $M_\odot$ yr$^{-1}$ kpc$^{-2}$, as is characteristic of so-called ``maximum starbursts" (eg. Scoville \hyperlink{S03}{2003}, Thompson et al. \hyperlink{T05}{2005}).
	
	CRLE is likely in a protocluster environment (see Section 5 of P18 for evidence of a galaxy overdensity around CRLE). Indeed its identification was made possible given its close proximity to the gas-rich main-sequence galaxy HZ10 at $z=5.654$ (J2000 $10^\mathrm{h}00^\mathrm{m}59^\mathrm{s}.3$, $1^\mathrm{o}33'19''.6$, Capak et al. \hyperlink{C}{2015}, Pavesi et al. \hyperlink{Pav}{2016}), which is at a distance of only 13$^{''}$ on the sky ($\sim$77 kpc) and was the main target of the observations resulting in the discovery of CRLE. 
	
	Here we expand on the results of P18 by making further measurements of CO(5$\rightarrow$4) and CO(6$\rightarrow$5) line emission in both CRLE and HZ10. There are very few $z > 5$ sources available for the study of higher-$J$ CO lines. Thus, the value of the observations in this work are two-fold: we place physical constraints on the conditions in the star-forming regions of CRLE and HZ10 and we also add to a very limited database of higher-$J$ CO line observations at $z > 5$. The latter allows for a general improvement in our understanding of the star-forming conditions present in these high-redshift galaxies.
	
	The paper is organized as follows: in Section 2 we describe the observations that led to the results presented in this work. In Section 3 we show the line emission properties and compare our results for the CO(5$\rightarrow$4) and CO(6$\rightarrow$5) transitions to the results found by P18 for the CO(2$\rightarrow$1) transition. In Section 4 we describe the modeling carried out in this work, using both large velocity gradient (LVG) and spectral energy distribution (SED) analysis. In Section 5 we compare our findings to previous work and discuss their impact in more detail. In this work we adopt a flat $\Lambda$CDM cosmology with $H_0$ = 70 km s$^{-1}$ Mpc$^{-1}$ and $\Omega_M$ = 0.3.
	
	\section{Observations}
	
	We observed the CO(5$\rightarrow$4) and CO(6$\rightarrow$5) transitions, which are shifted to $86.447 \pm 0.008$ and $103.723 \pm 0.008$ GHz at the redshift of $z = 5.6664 \pm 0.0004$ for our main target CRLE and to $86.604 \pm 0.008$ and $103.912 \pm 0.008$ GHz at $z = 5.6543 \pm 0.0004$ for HZ10. 
	
	Observations were carried out on 2018 November 17 and 2019 February 25 utilizing the Northern Extended Millimeter Array (NOEMA). The total on-source time was 3.5 hours and 10 antennas were used in the C configuration. The pointing was centered on CRLE. Both lines were observed at the same time by placing them in the lower and upper sidebands, respectively.
	
	The BL Lac type object 3C273 was used for RF bandpass and flux calibration while J0948+003 was used for phase and amplitude calibration. MWC349 was also used for flux calibration.
	
	The correlator was set up with 8 units, each having a width of 4064 MHz and a resolution of 2 MHz. The 8 units were arranged in a dual polarization set-up with 7.679 GHz wide sidebands\footnote{Note that the total bandwidth is less than 8.128 GHz as the edges of each polarization unit are flagged; there are two units per sideband}.
	
	Data calibration was carried out using \texttt{GILDAS}\footnote{https://www.iram.fr/IRAMFR/GILDAS}.
	
	The synthesized beam size for imaging of the CO(5$\rightarrow$4) data is 4$.^{''}$4 $\times$ 1$.^{''}$6 while for the CO(6$\rightarrow$5) data it is 3$.^{''}$7 $\times$ 1$.^{''}$5 when adopting natural baseline weighting. The rms noise in the phase center is 68 $\mu$Jy beam$^{-1}$ for the lower sideband in a 69 km s$^{-1}$ (20 MHz) wide channel and 95 $\mu$Jy beam$^{-1}$ for the upper sideband in a 57 km s$^{-1}$ (20 MHz) wide channel.
	
	\begin{figure*}
		\centering
		\subfigure{\includegraphics[width=170mm]{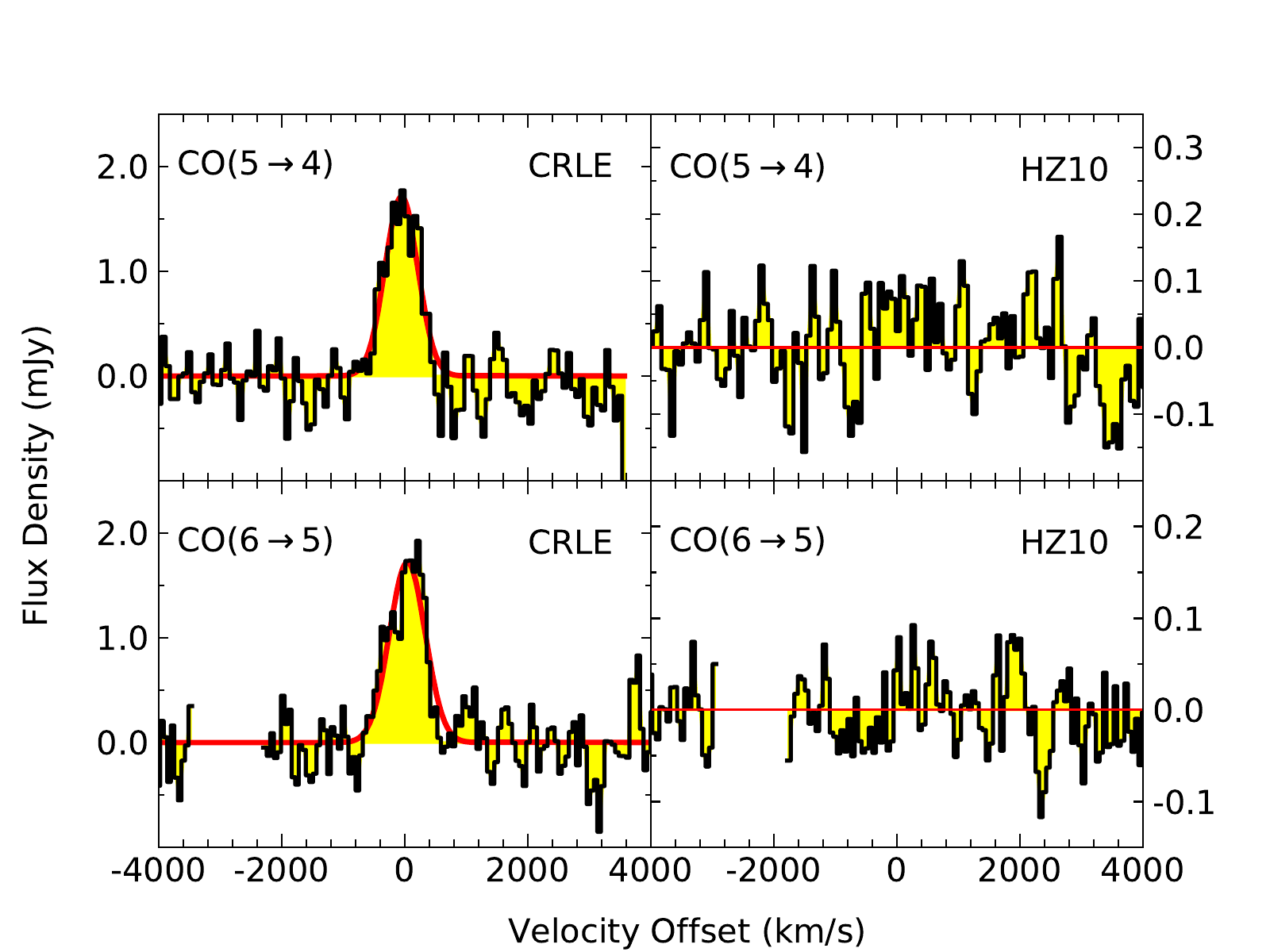}}
		\caption{The CO(5$\rightarrow$4) and CO(6$\rightarrow$5) continuum-subtracted line spectra (histograms). Gaussian fits to the line emission are shown as red curves. Mean redshifts of $z = 5.667$ and $z = 5.654$ were adopted as the velocity reference for CRLE and HZ10, respectively. No continuum was subtracted for HZ10 since none was detected. The channel widths are 69 and 58 km s$^{-1}$ for CO(5$\rightarrow$4) and CO(6$\rightarrow$5), respectively. The spectra were extracted from the peak pixel in the image cube, since the emission remains unresolved in our data. The gap in the CO(6$\rightarrow$5) spectrum is due to the flagging of edge channels in each of the two polarization units in the sideband.}
	\end{figure*}
	
	\begin{figure*}
		\centering
		\subfigure{\includegraphics[width=170mm]{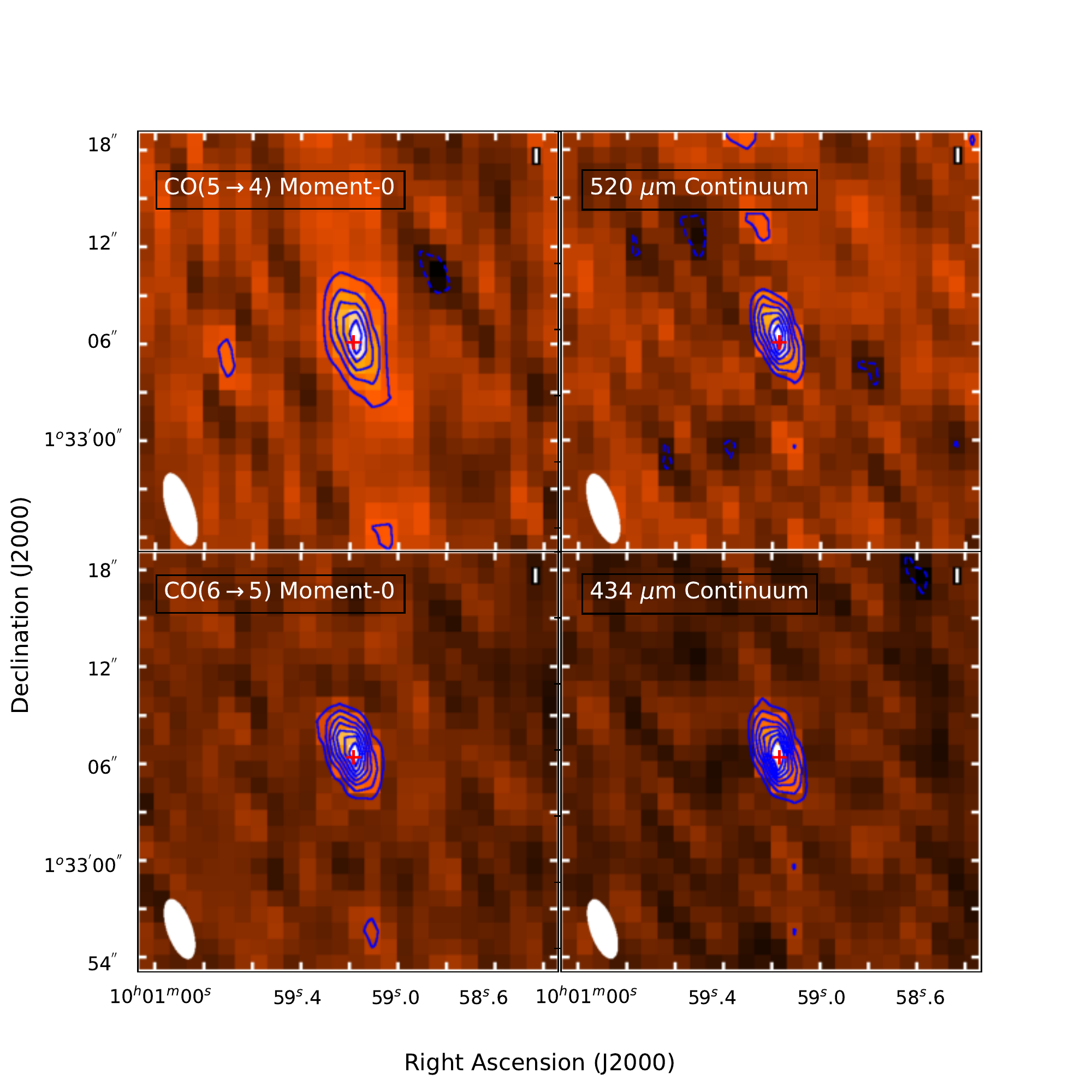}}
		\caption{Moment-0 and continuum maps for the sidebands in which CO(5$\rightarrow$4) (upper) and CO(6$\rightarrow5$) (lower) emission was observed for CRLE. Contours are multiples of $\pm 2\sigma$. The rms values for the moment-0 and continuum maps are 68 and 21 $\mu$Jy beam$^{-1}$ for CO(5$\rightarrow$4) and 95 and 32 $\mu$Jy beam$^{-1}$ for CO(6$\rightarrow$5). For the moment-0 maps we use velocity widths of 1518 and 1254 km s$^{-1}$ for CO(5$\rightarrow$4) and CO(6$\rightarrow5$) respectively. Red crosses denote the position of peak emission in each panel. The beam is shown as a filled ellipse in the bottom left corner.}
	\end{figure*}
	
	\begin{figure*}
		\centering
		\subfigure{\includegraphics[width=170mm]{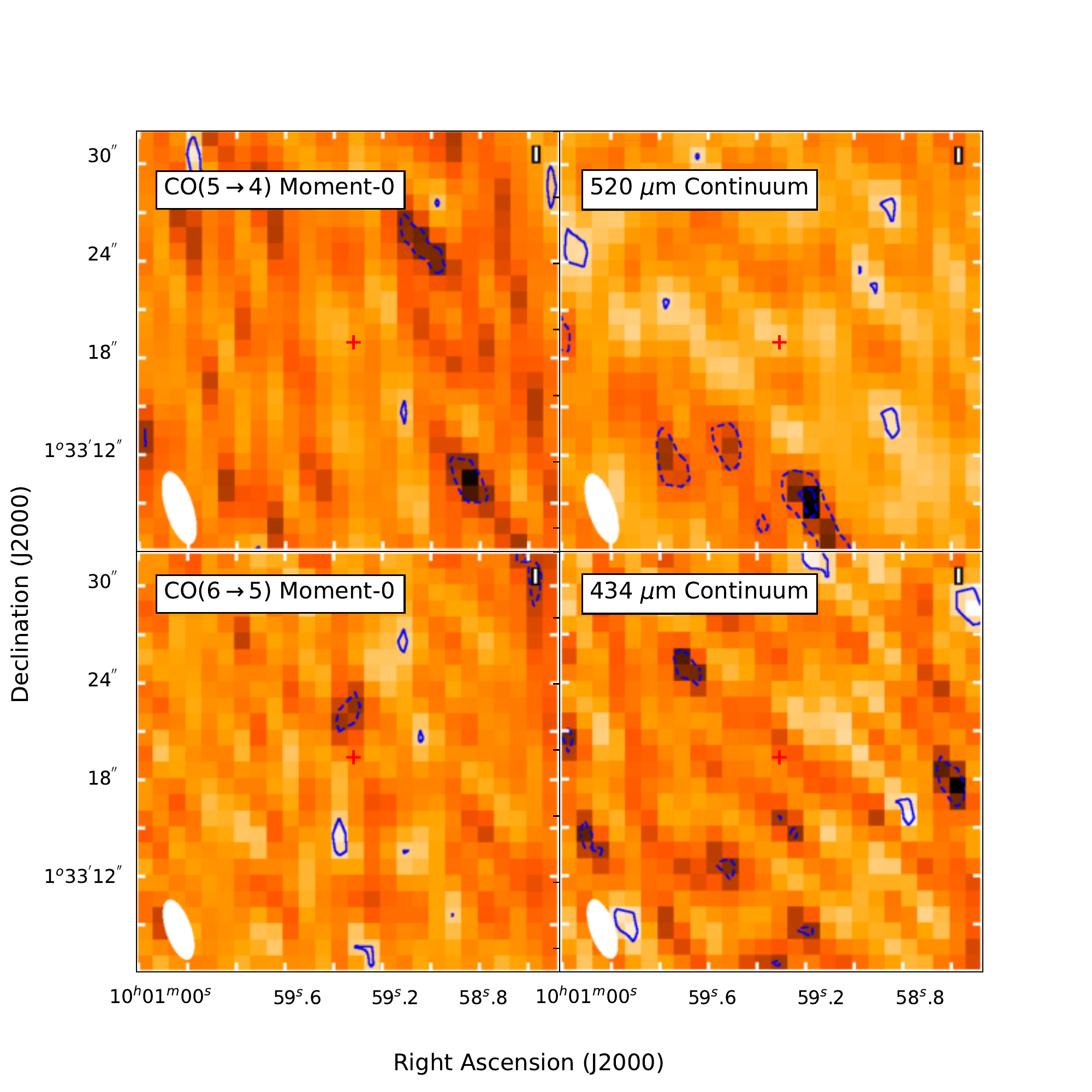}}
		\caption{Moment-0 and continuum maps for the sidebands in which CO(5$\rightarrow$4) (upper) and CO(6$\rightarrow5$) (lower) emission was observed for HZ10. Contours are multiples of $\pm 2\sigma$. The rms values for the moment-0 and continuum maps are 88 and 27 $\mu$Jy beam$^{-1}$ for CO(5$\rightarrow$4) and 99 and 31 $\mu$Jy beam$^{-1}$ for CO(6$\rightarrow$5). For the moment-0 maps we use a velocity width of 1311 km s$^{-1}$ for both CO(5$\rightarrow$4) and CO(6$\rightarrow5$). Red crosses (fixed to the same position in all 4 panels) denote the position of the [CII] emission peak in HZ10 (Capak \hyperlink{C}{2015}). The beam is shown as a filled ellipse in the bottom left corner.}
	\end{figure*}	
	
	\begin{figure*}
		\centering
		\subfigure{\includegraphics[width=170mm]{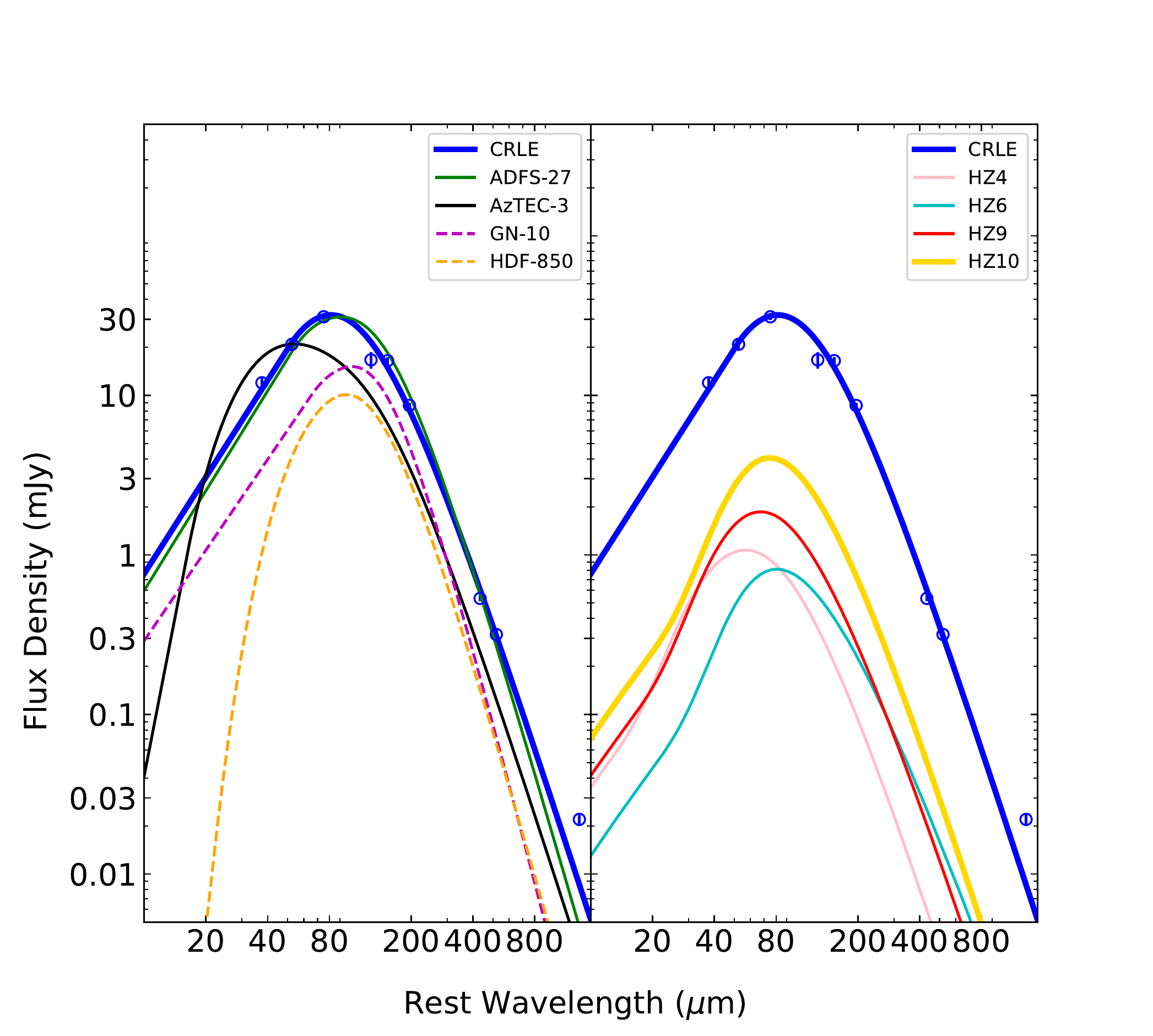}}
		\caption{Dust SED fits for five SMGs (left) and four MS galaxies (right) all at $z>5$, including CRLE (blue lines and data points) in both panels. The comparison sample shows the difference in the SED between HyLIRGs (CRLE, ADFS-27, AzTEC-3) and lower luminosity galaxies (GN10, HDF850.1) in the left panel. The right panel compares CRLE with the four MS galaxies. SED fits for ADFS-27 are from Riechers et al. (\hyperlink{R3}{2017}); for AzTEC-3, GN10 and HDF850.1 from Riechers et al. (\hyperlink{R4}{2020}); and for all HZ galaxies from Faisst et al. (\hyperlink{Fa}{2020}).}
	\end{figure*}
	
	\begin{figure}
		\centering
		\subfigure{\includegraphics[width=100mm]{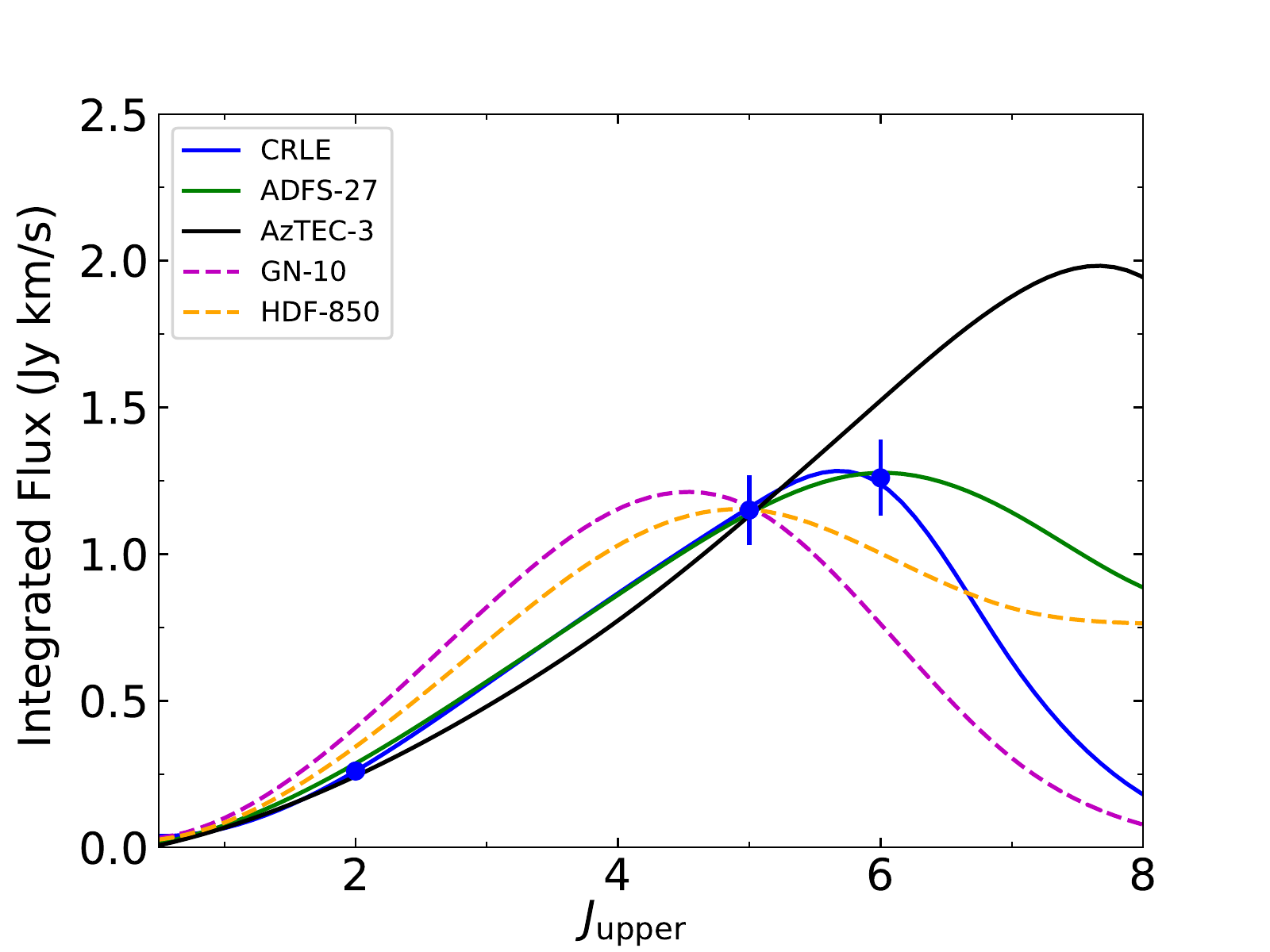}}
		\caption{The best fitting CO excitation line ladders for various SMGs at $z > 5$. CRLE, ADFS-27 (Riechers et al. \hyperlink{R3}{2017}) and AzTEC-3 (Riechers et al. \hyperlink{R4}{2020}) are HyLIRGs (solid lines). GN10 and HDF850.1 (Riechers et al. \hyperlink{R4}{2020}) have lower luminosity (dashed lines). Values of the flux are normalized to the CO(5$\rightarrow$4) flux of CRLE. For CRLE: $T_{\mathrm{kin}}$ = 30 K, filling factor = 0.3, n(H$_2$)=10$^{4.0}$ cm$^{-3}$. The CO(2$\rightarrow$1) measurement is from P18.}
	\end{figure}
	
	\section{Line and Continuum Emission}
	
	\subsection{CRLE}
	
	We detect CO(5$\rightarrow$4) and CO(6$\rightarrow$5) line emission and the adjacent continuum in CRLE. We use tasks in the Common Astronomy Software Applications (\texttt{CASA}) package to obtain the results summarized in Table 1. To obtain line emission properties we extract spectra from our image cube by selecting the peak pixel and using the Spectral Profile tool in \texttt{CASA}, after verifying that the emission remains spatially unresolved in the moment-0 maps. We fit Gaussian functions to the spectral profiles after the subtraction of continuum emission. The fits are shown in Figure 1 with the continuum subtracted using the \texttt{CASA} task \texttt{uvcontsub}. From the fits we obtain the peak flux density, full width at half maximum (FWHM) and integrated line intensity for each transition. We also extract line intensities from the moment-0 maps, finding consistent results.
	
	Averaging the redshift found for each of the two transitions, we find $z = 5.6664 \pm 0.0004$, in agreement with the value of $z = 5.6666 \pm 0.0008$ found by P18. 
	
	We calculate a ratio of line luminosities to the value of ($7.0 \pm 0.5$) $\times$ 10$^{10}$ K km s$^{-1}$ pc$^2$ obtained by P18 for the CO(2$\rightarrow$1) transition finding $L'_{5\rightarrow4}/L'_{2\rightarrow1}$ = 0.7 $\pm$ 0.1 and $L'_{6\rightarrow5}/L'_{2\rightarrow1}$ = 0.5 $\pm$ 0.1. 
	
	We obtain continuum fluxes by fitting a 2D Gaussian to the emission at the position of CRLE from the line-free channels using the \texttt{CASA} task \texttt{imfit}. We confirm the results by calculating the flux density with the \texttt{CASA} task \texttt{specflux} for a continuum image obtained from \texttt{imcontsub}.
	
	\subsection{HZ10}
	
	We do not detect CO(5$\rightarrow$4) or CO($6\rightarrow$5) line emission at the position of HZ10 at a signal-to-noise ratio of $>3$. Thus, we place an upper limit on the line emission properties from the spectrum of HZ10 (see Table 1). We do this by finding the integrated flux density in the moment-0 map (taken over the CO(2$\rightarrow$1) linewidth, centered on the CO(2$\rightarrow$1) redshift; see Figure 3) and adding 3 times the rms noise. We then use the upper limits on the CO(5$\rightarrow$4) and CO(6$\rightarrow$5) line intensities to derive upper limits on the respective line luminosities, finding $L'_{5\rightarrow4} < 0.76$ and $L'_{6\rightarrow5} < 0.60$ $\times$ 10$^{10}$ K km s$^{-1}$ pc$^2$. We also derive upper limits on the ratio of the line luminosities to the detected CO(2$\rightarrow$1) line luminosity in Pavesi et al. (\hyperlink{Pa}{2019}) finding $L'_{5\rightarrow4}/L'_{2\rightarrow1} < 0.26$ and $L'_{6\rightarrow5}/L'_{2\rightarrow1} < 0.21$.
	
	We do not detect continuum emission at the position of HZ10 and the upper limits for each sideband are reported in Table 1. 
	
	\section{Modeling}
	
	\subsection{Large Velocity Gradient}
	
	We model the line emission in CRLE by using a large velocity gradient (LVG) code to calculate the Rayleigh-Jeans temperature for CO transitions from $J=1\rightarrow 0$ to $J=11\rightarrow 10$ on a grid of kinetic temperature ($T_{\mathrm{kin}}$) and density. The grid spans $T_{\mathrm{kin}}$ from 5-100 K and H$_2$ densities from 10$^{2}$-10$^{6.1}$ cm$^{-3}$. We assume a CO-H$_2$ abundance ratio of 10$^{-5}$ (see Weiss et al. \hyperlink{W05}{2005}) and calculate the CMB temperature as $T_{\rm{CMB}} = 2.73(1+z) = 18.2$ K (at $z=5.667$). 
	
	We take the physical size of the emission region to be the size of the [CII] emission reported in P18: (3.7 $\pm$ 0.2) $\times$ (2.4 $\pm$ 0.3) kpc$^2$. Our value for the filling factor is in relation to this area.
	
	We implement an MCMC code making use of the python package \texttt{emcee} (Foreman-Mackey et al. \hyperlink{FM}{2013}) to place statistical limits on our findings. We find that our results for the CO(5$\rightarrow$4) and CO(6$\rightarrow$5) transitions are reproduced for kinetic temperatures in the range $T_{\rm{kin}}$ $\sim$ 25-70 K; for H$_{2}$ densities in the range $n$(H$_2$) $\sim$ 10$^{3.8}$-10$^{4.1}$ cm$^{-3}$; and for filling factor in the range $\sim$ 0.1-0.5 (see Appendix). This suggests a fairly dense gas environment in regions traced by the CO(5$\rightarrow$4) and CO(6$\rightarrow$5) transitions, likely associated with high star formation activity. 
	
	\subsection{Spectral Energy Distribution}
	
	We model the spectral energy distribution (SED) of CRLE by fitting its continuum data with a modified blackbody smoothly connected to a mid-IR power law\footnote{CRLE is located behind a local foreground spiral galaxy, however the contribution of this galaxy to the continuum emission at the wavelengths we consider is expected to be negligible (see P18)} (see Figure 4). The model fitting uses the \texttt{mbb\_emcee}\footnote{https://github.com/aconley/mbb\_emcee} code (Riechers et al. \hyperlink{R2}{2013}, Dowell et al. \hyperlink{Do}{2014}) wrapped in the affine invariant MCMC package described in Foreman-Mackey et al. (\hyperlink{FM}{2013}). We fit for the dust temperature $T_{\rm{d}}$, the rest-frame wavelength at which the optical depth becomes unity $\lambda_0$, a mid-IR power law index $\alpha$ and the dust emissivity parameter $\beta$, using uniform, nonconstraining priors (see Table 2 for results).
	
	We find a dust temperature of $T_{\rm{d}}$ = 57$\pm$4 K and an IR luminosity of $L_{\mathrm{IR}}$ = $(3.0\pm 0.1) \times 10^{13} L_{\odot}$. The IR luminosity is obtained by integrating between 8 and 1000 $\mu$m. By integrating between 42.5 and 122.5 $\mu$m we obtain the FIR luminosity, finding $L_{\mathrm{FIR}}$ = $(1.7 \pm 0.1) \times 10^{13} L_{\odot}$.
	
	It is worth noting that if we were to keep the kinetic temperature in our LVG model fixed at the dust temperature found from our SED fit, we would find an H$_2$ density of $\sim$ 10$^{3.9}$ cm$^{-3}$, agreeing with the optimal range of values found by the MCMC code.
	
	Interestingly the 1.3 mm continuum measurement (22 $\pm$ 2 $\mu$Jy; P18) lies slightly above the best-fitting curve of our SED model. Together with a known 3 GHz ($\lambda_{\rm{rest}}$ = 15 mm) continuum measurement (24.3 $\pm$ 3.8 $\mu$Jy; Smol{\^c}i{\'c} et al. \hyperlink{S17}{2017}), this may be an indication that free-free emission is contributing at this wavelength.
	
	\subsection{Gas Mass}
	
	The conversion of higher $J$ CO line intensities to a CO(1$\rightarrow$0) line intensity has been the subject of much debate (Geach \& Papadopoulos \hyperlink{GP}{2012}, Carilli \& Walter \hyperlink{CW}{2013}, Narayanan \& Krumholz \hyperlink{N}{2014}). Here we calculate the CO(1$\rightarrow$0) line intensity in two different ways: by taking the value from our LVG model (see Figure 5) and by following the conversion presented in Narayanan \& Krumholz (\hyperlink{N}{2014}). We then convert the CO(1$\rightarrow$0) intensity to a total molecular gas mass, thus obtaining two separate estimates for the gas mass.
	
	To convert the CO(1$\rightarrow$0) line intensities to total molecular gas masses we convert them to line luminosities and then use a conservative estimate for the final conversion factor to total molecular gas mass $\alpha_{\mathrm{CO}}$ of 1 $M_\odot$ (K km s$^{-1}$ pc$^2$)$^{-1}$. We use this estimate because it is consistent with the gas-to-dust mass ratio range of $\sim$ 35-75 for CRLE (P18).
	
	Our primary estimate for the total molecular gas mass is obtained from our LVG model. We find a total mass of $M_{\mathrm{gas}} = (7.1 \pm 1.3) \times 10^{10}$ $M_\odot$. This value agrees well with the one found in P18 ($(7.0 \pm 0.5) \times 10^{10}$ $M_\odot$) by the direct conversion of $L'_{2\rightarrow1}$ to molecular gas mass.
	
	The method of Narayanan \& Krumholz (\hyperlink{N}{2014}) gives a second estimate of $M_{\mathrm{gas}}$. We first calculate the star formation rate surface density $\Sigma_{\mathrm{SFR}}$ from the measured source size and the IR luminosity. Based on the FIR continuum emission we derive a source size of (23.0 $\pm$ 4.6) kpc$^2$. Combining this with our measured value for $L_{\mathrm{IR}}$ of $(3.0 \pm 0.1) \times 10^{13} L_{\odot}$ (and accounting for the fact that the source size is taken as the FWHM of a Gaussian) yields an IR surface density of $\Sigma_{\mathrm{IR}} = (0.97 \pm 0.23) \times 10^{12} L_\odot$ kpc$^{-2}$. This leads to an estimate for $\Sigma_{\mathrm{SFR}}$ of $\sim 97 \pm 23$ $M_\odot$ yr$^{-1}$ kpc$^{-2}$ (Kennicutt \hyperlink{K}{1998}). We then apply equation (19) (using Table 3 for the best-fit values of $A, B, C$) in Narayanan \& Krumholz (\hyperlink{N}{2014}). This method gives a value of $M_{\mathrm{gas}} = (8.5 \pm 2.6) \times 10^{10}$ $M_\odot$, about 20$\pm$35$\%$ higher than the value found above but fully consistent within the uncertainties.
	
	Finally, we may also estimate the ISM mass of CRLE using equation (14) in Scoville et al. (\hyperlink{Sco}{2017}). This method uses the long-wavelength dust continuum as a quantitative probe. Using the continuum flux at 520 $\mu$m we derive an estimate for the total ISM mass of $(4 \pm 1) \times 10^{11}$ $M_\odot$ with this technique. In P18 rest-frame 1.3 mm observations were used with the same method to estimate a total ISM mass of $8.0 \times 10^{11}$ $M_\odot$. The discrepancy may be due to free-free emission contributing to a higher continuum flux measurement at 1.3 mm.
	
	\section{Discussion}
	
	\subsection{Gas Excitation}
	
	The ratios of the line luminosities give insight into the physical conditions present in the gas. The relatively high ratios of $L'_{5\rightarrow4}$/$L'_{2\rightarrow1}$ and $L'_{6\rightarrow5}$/$L'_{2\rightarrow1}$ could be partially explained by the higher temperature of the CMB at $z=5.667$ ($T_{\mathrm{CMB}} = 18.2$ K) helping to populate higher rotational levels\footnote{A rough estimate based on the method described by da Cunha et al. (\hyperlink{dC}{2013}) yields a value for $L'_{5\rightarrow4}$/$L'_{2\rightarrow1}$ about a factor of two lower for an equivalent galaxy at $z=0$.} (da Cunha et al. \hyperlink{dC}{2013}). The CMB also suppresses emission from gas at low brightness temperatures as our measurements are taken in contrast to it. The ratio of $L'_{6\rightarrow5}$/$L'_{5\rightarrow4}$ is less than unity (0.8 $\pm$ 0.2), suggesting that the gas traced by these lines is subthermally excited. 
	
	We may also compare the line luminosities to other DSFGs\footnote{We generalize all high-$z$ galaxies selected at infrared or sub-mm wavelengths as dusty star-forming galaxies (DSFGs)} at similar redshifts. Riechers et al. (\hyperlink{R4}{2020}) compute $L'_{\mathrm{line}}$ for 3 sources at $z>5$: GN10, AzTEC-3 and HDF850.1. Our value of 0.7$\pm$0.1 for $L'_{5\rightarrow4}$/$L'_{2\rightarrow1}$ falls between the values of $\sim$0.5 for GN10 and HDF850.1 and the value of $\sim$0.8 for AzTEC-3. For $L'_{6\rightarrow5}$/$L'_{2\rightarrow1}$ our value of 0.5$\pm$0.1 again
	falls between the lower values of $\sim$0.2 and $\sim$0.3 for GN10 and HDF850.1 and the much higher value of $\sim$0.8 for AzTEC-3. Thus, the ratios seem fairly average for $z>5$ DSFGs.
	
	This is further corroborated by comparing the line ratios to the average value obtained by Spilker et al. (\hyperlink{Sp}{2014}) by stacking 22 high-redshift (spanning $z = 2.0-5.7$) DSFGs discovered by the South Pole Telescope (SPT). Of the 22 galaxies, 10 have both CO(2$\rightarrow$1) and CO(5$\rightarrow$4) measurements (see Weiss et al. \hyperlink{We}{2013} and Aravena et al. \hyperlink{Ar}{2016}). For these 10 galaxies, Spilker et al. find $L'_{5\rightarrow4}$/$L'_{2\rightarrow1} = 0.61 \pm 0.04$. Of the 10 galaxies, two have a CO(6$\rightarrow$5) measurement and for these two one can compare the results of Spilker et al. (\hyperlink{Sp}{2014}) and Aravena et al. (\hyperlink{Ar}{2016}) to find  $L'_{6\rightarrow5}$/$L'_{2\rightarrow1} = 0.36 \pm 0.05$. The ratios are similar to those found for CRLE.
	
	Considering the observed H$_{2}$ density range, the kinetic temperature range is typical for excitation of $J=5$ and $J=6$ lines in a high-redshift galaxy, as can be seen by comparing our results to the Spectral Line Energy Distributions (SLEDs) in Weiss et al. (\hyperlink{W07}{2007}).
	
	Comparing to the average CO excitation in a sample of 40 luminous SMGs conducted by Bothwell et al. (\hyperlink{Bo}{2013}) we observe higher CO excitation in CRLE (see Figure 9 in Apostolovski et al. \hyperlink{A}{2019}). This comes with the important caveat that in Bothwell et al. (\hyperlink{Bo}{2013}), $L'_{5\rightarrow4}$ and $L'_{6\rightarrow5}$ are only known for the source GN20, for which $L'_{5\rightarrow4}/L'_{2\rightarrow1} = 0.55 \pm 0.31$ and $L'_{6\rightarrow5}/L'_{2\rightarrow1} = 0.31 \pm 0.11$ (Carilli et al. \hyperlink{C10}{2010}, Hodge et al. \hyperlink{H12}{2012}).
	
	\subsection{Dust Temperature and FIR Luminosity}
	
	The dust temperature\footnote{Note that we refer to the SED dust temperature, not to be confused with the peak dust temperature (see Casey et al. \hyperlink{Cas}{2012})} can be compared to the work of Cortzen et al. (\hyperlink{Co}{2020}), which compiles a number of datasets to produce a plot of $T_{\rm{d}}$ as a function of $z$. Assuming the linear trend of Schreiber et al. (\hyperlink{Sc}{2018}) holds for $z > 5$, one can estimate a dust temperature of $T_{\rm{d}} \sim 55$ K for a MS galaxy at the redshift of CRLE. While this appears to be a good match to our data, it is worth noting that there is some debate whether a $T_{\rm{d}}-z$ relation such as that given by Schreiber et al. (\hyperlink{Sc}{2018}) truly exists. For example, Dudzeviciute et al. (\hyperlink{Dz}{2020}) suggest that the evolution of $T_{\rm{d}}$ with $z$ disappears after correcting for observational biases (see also discussion by Riechers et al. \hyperlink{R4}{2020}). Faisst et al. (\hyperlink{Fa}{2020}) match FIR luminosity across bins in redshift and report an increase in $T_{\rm{d}}$ up to $z \sim 4$, with a flatter curve at higher redshifts.
	
	For starburst galaxies at high redshifts ($z > 2.5$), lower dust temperatures have been suggested than for main sequence galaxies at similar redshifts (see for example Bethermin et al. \hyperlink{B}{2015}, Jin et al. \hyperlink{J}{2019}). In Cortzen et al. (\hyperlink{Co}{2020}) the argument is made that the dust temperature appears artificially low due to the assumption of optically thin far-infrared dust emission. This argument is consistent with the findings of several previous studies indicating that the dust in high-redshift massive starbursts could remain optically thick out to rest-frame wavelengths of $\lambda_0$ = 100-200 $\mu$m (Riechers et al. \hyperlink{R2}{2013}, Spilker et al. \hyperlink{Sp2}{2016}, Hodge et al. \hyperlink{H}{2016}). Accounting for the effects of optical depth leads to higher values of $T_{\rm{d}}$ and removes the tension of a large difference between the measured excitation temperature $T_{\mathrm{ex}}$ and $T_{\rm{d}}$. 
	
	As an example, in Jin et al. (\hyperlink{J}{2019}) it is noted that an optically thin fit for the galaxy ID 85001929 finds a dust temperature of $T_{\rm{d}} = 42 \pm 3$ K while an optically thick fit finds $T_{\rm{d}} = 61 \pm 8$ K. Riechers et al. (\hyperlink{R4}{2020}) study the same source under the assumption of an optically thick medium and find $T_{\rm{d}} = 59.0^{+7.7}_{-16.7}$ K. 
	
	The value of $T_{\rm{d}}$ reported in this work is significantly higher than that reported by P18 (41$^{+6}_{-2}$ K). P18 finds a value for $\lambda_0$, the wavelength at which the optical depth becomes unity, of 16$^{+64}_{-14}$ $\mu$m, while our model finds a value of 147$^{+17}_{-22}$ $\mu$m. By the arguments outlined above, the higher value of $\lambda_0$ leads naturally to a higher value for $T_{\rm{d}}$.
	
	It is worth noting that this dust temperature obtained from our SED fit is not the same as the dust temperature used to estimate the ISM mass in the method of Scoville et al. (\hyperlink{Sco}{2017}). The SED fit captures mainly smaller dust grains while the method of Scoville et al. (\hyperlink{Sco}{2017}) uses a ``standard" dust temperature of 25 K meant to capture primarily the larger cold dust grains. This may be part of the reason that the method overestimates the ISM mass, as the CMB temperature at $z=5.7$ of 18.2 K is quite high in comparison to the standard 25 K temperature.
	
	One can further compare the dust temperature found in this work to the 17 $z>5$ DSFGs compiled by Riechers et al. (\hyperlink{R4}{2020}). CRLE shows a fairly high dust temperature, similar to galaxies such as HLock-102 (Dowell et al. \hyperlink{Do}{2014}), ADFS-27 (Riechers et al. \hyperlink{R3}{2017}) and the optically thick estimate for ID 85001929 (Jin et al. \hyperlink{J}{2019}). 
	
	The dust temperature is consistent (within the errors) with the mean kinetic temperature from the MCMC analysis. Based on our current observations, no external source of mechanical energy input in the ISM appears to be required. However, such effects may only become visible in transitions higher than CO(6$\rightarrow$5). 
	
	Based on constraints on the IR luminosity function at $z>5$, it is clear that the 17 $z>5$ DSFGs (of which 7 are HyLIRGs) in Riechers et al. (\hyperlink{R4}{2020}) represent only a small fraction of the total number of DSFGs that exist at these redshifts (Gruppioni et al. \hyperlink{G}{2020}). The currently detected sample represents the bright end of the luminosity function: models suggest that the fraction of HyLIRGs among all $z>5$ DSFGs is much lower than the fraction observed in the sample of Riechers et al. (\hyperlink{R4}{2020}).
	
	In this sense, though the far-infrared luminosity of CRLE may not appear to be exceptional based on a comparison to the sample of Riechers et al. (\hyperlink{R4}{2020}), its value is likely much higher than in an average $z>5$ DSFG due to the luminosity function constraints. 
	
	It is interesting to analyze CRLE in the context of the other 6 HyLIRGs in the sample of Riechers et al. (\hyperlink{R4}{2020}). The average gas mass among the HyLIRGs is $1.6 \times 10^{11} M_\odot$ and the average SFR is $2700$ $M_\odot$ yr$^{-1}$. These numbers can be compared to the slightly lower gas mass and SFR of CRLE: $7.1 \times 10^{10} M_\odot$ and $2200$ $M_\odot$ yr$^{-1}$, respectively. 
	
	While CRLE has the second-lowest values of both $M_{\mathrm{gas}}$ and SFR among the 7 HyLIRGs, it has the fourth-highest value for the ratio $M_{\mathrm{gas}}$/SFR, indicating that its star-forming properties are likely fairly average in comparison with other $z>5$ HyLIRGs.
	
	Looking at both the values of $M_{\mathrm{gas}}$ and $M_{\mathrm{dust}}$ ($(1.3 \pm 0.3) \times 10^9 M_\odot$, P18) shows that they are comparable most specifically to those of GN10 ($M_{\mathrm{gas}} = (7.1 \pm 0.9) \times 10^{10} M_\odot$, $M_{\mathrm{dust}} = (1.1 \pm 0.4) \times 10^9 M_\odot$, Riechers et al. \hyperlink{R4}{2020}). In this sense CRLE is perhaps quite similar to GN10 (the two galaxies also share, within uncertainties, close values of the dust temperature $T_{\rm{d}}$).
	
	\subsection{Constraints on HZ10}
	
	The limits of $L'_{5\rightarrow 4} < 0.76$ and $L'_{6\rightarrow 5} < 0.60$ $\times$ 10$^{10}$ K km s$^{-1}$ pc$^2$ can be compared to results from the ALMA Spectroscopic Survey (Boogaard et al. \hyperlink{Boo}{2020}). The survey finds an average line-luminosity of $L'_{5\rightarrow 4} = 0.32 \pm 0.05$ $\times$ 10$^{10}$ K kms$^{-1}$ pc$^2$ for three sources at redshifts within the range $z = [1.32,1.55]$; and $L'_{6\rightarrow 5} = 0.27 \pm 0.08$ $\times$ 10$^{10}$ K km s$^{-1}$ pc$^2$ for three sources at redshifts within $z = [1.55,2.00]$. The average ratio of $L'_{5\rightarrow 4}/L'_{2\rightarrow 1} = 0.25 \pm 0.04$ may be similar to that of HZ10 considering its upper limit of $< 0.26$.
	
	Daddi et al. (\hyperlink{Da}{2015}) study four galaxies at redshifts within the range $z = [1.41,1.53]$. They find an average line luminosity of $L'_{5\rightarrow 4} = 0.53 \pm 0.07$ $\times$ 10$^{10}$ K kms$^{-1}$ pc$^2$ and a ratio of $L'_{5\rightarrow 4}/L'_{2\rightarrow 1} = 0.30 \pm 0.03$.
	
	From these results we see that the limits on HZ10 may imply a comparable ratio of $L'_{5\rightarrow 4}/L'_{2\rightarrow 1}$ to the average value of MS galaxies at lower redshift, suggesting similar gas excitation. As the higher CMB temperature at $z>5$ leads naturally to the excitation of higher energy levels the results may imply the presence of a cold gas reservoir in HZ10 of equal or larger size than the average for MS galaxies at lower redshift.
	
	Altogether the comparatively low gas excitation indicated by our constraints solidifies previous estimates of the galaxy's gas mass and star formation efficiency. Pavesi et al. (\hyperlink{Pa}{2019}) find a gas mass of (1.2 $\pm$ 0.5) $\times$ 10$^{11}$ $M_\odot$ and a gas depletion timescale of 960$^{+1200}_{-470}$ Myr; these results imply that HZ10 is very rich in molecular gas. A CO luminosity to total molecular gas mass conversion factor $\alpha_{\rm{CO}}$ of 4.2$^{+2}_{-1.7}$ $M_\odot$ (K km s$^{-1}$ pc$^2$)$^{-1}$ is also derived in Pavesi et al. (\hyperlink{Pa}{2019}), a value compatible with $z \approx 1.5$ MS disk galaxies. Furthermore, the star formation efficiency is comparable to lower-redshift MS galaxies. The low excitation indicated by the constraints in this work thus supports the finding that HZ10 is a gas-rich MS galaxy that may have similar properties to lower-redshift disk galaxies.
	
	\section{Conclusions}
	
	We study the CO(5$\rightarrow$4) and CO(6$\rightarrow$5) transitions in CRLE, the most IR luminous galaxy at $z>5$ in the COSMOS field; and in HZ10, the most gas-rich MS galaxy currently known at $z>5$. We compare line luminosities to those observed for the CO(2$\rightarrow$1) transition, finding ratios consistent with other high-redshift DSFGs for CRLE; and consistent with MS galaxies at $z\approx 1.5$ for HZ10. 
	
	We model the line emission in CRLE using an LVG code and find evidence for a high-density environment traced by the two observed CO lines. The CO excitation ladder shows higher excitation than in average DSFGs; however the excitation is similar to other $z>5$ DSFGs, and comparable to the SPT average (based on 10 direct measurements of CO(5$\rightarrow$4) and 2 of CO(6$\rightarrow$5)).
	
	We find the dust temperature and IR luminosity using a spectral energy distribution fit to continuum data. The value for $T_{\rm{d}}$ of $58^{+2}_{-4}$ K is similar to that of other $z>5$ DSFGs.
	
	We calculate the total molecular gas mass, finding a similar value to the gas mass of GN10. We also find that the IR luminosity is higher than in most $z > 5$ DSFGs. 
	
	The properties of individual galaxies such as CRLE are very important in studying star-forming conditions at $z>5$: by studying their CO line luminosities we can begin to answer questions regarding the contribution of these HyLIRGs to the total molecular gas content and thus the total ``fuel" for star formation in the early Universe (see Riechers et al. \hyperlink{R5}{2019}, Popping et al. \hyperlink{Pop}{2016}, Vallini et al. \hyperlink{V}{2016}).
	
	We constrain the physical conditions in HZ10, now the only main-sequence galaxy with constraints on CO excitation at this redshift. We find a comparable ratio of $L'_{5\rightarrow4}/L'_{2\rightarrow1}$ to average MS galaxies at $z\approx 1.5$, suggesting that similar galaxies may already have existed only one billion years after the Big Bang. 
	\\
	\\
	\\
	D.V. and D.R. acknowledge support from the National Science Foundation under grant numbers AST- 1614213 and AST-1910107. D.R. also acknowledges support from the Alexander von Humboldt Foundation through a Humboldt Research Fellowship for Experienced Researchers. This work is based on observations carried out under project number S18DK with the IRAM NOEMA Interferometer. IRAM is supported by INSU/CNRS (France), MPG (Germany) and IGN (Spain). The authors thank Ian Smail and James Geach for providing an updated flux measurement from the SCUBA-2 COSMOS survey and Christian Henkel for the original version of the LVG code.
	
	\clearpage
	
	\begin{turnpage}
		\begin{table}
			\begin{tabular}{p{4cm} p{4cm} p{4cm} p{4cm} p{4cm}}
				\multicolumn{5}{c}{\textbf{Table 1}} \\
				\multicolumn{5}{c}{Parameters for CRLE (left) and HZ10 (right)} \\
				\hline
				\hline
				& CO(5$\rightarrow$4) & CO(6$\rightarrow$5) & CO(5$\rightarrow$4) & CO(6$\rightarrow$5) \\
				\hline
				$z$ & 5.6662 $\pm$ 0.0006 & 5.6665 $\pm$ 0.0005 & (5.6543) & (5.6543) \\
				$\nu_{\mathrm{obs}}$ (GHz)   & 86.447 $\pm$ 0.008    & 103.723 $\pm$ 0.008 & (86.604) & (103.912)\\
				$S_{\mathrm{peak}}$ (mJy)  & 1.49 $\pm$ 0.15 & 1.63 $\pm$ 0.16 & $<$0.263 & $<$0.298 \\
				$S_{\mathrm{cont}}$ (mJy)  & 0.317 $\pm$ 0.025 & 0.534 $\pm$ 0.023 & $<$0.082 & $<$0.092 \\
				$\mathrm{FWHM}$ (km/s) & 655 $\pm$ 82 & 744 $\pm$ 93 & (650) & (650)\\
				$I$ (Jy km s$^{-1}$) & 1.15 $\pm$ 0.12 & 1.26 $\pm$ 0.13 & $<$0.17 & $<$0.19 \\
				\hline
				$L'$ (10$^{10}$ K km s$^{-1}$ pc$^{2}$) & 4.9 $\pm$ 0.5 & 3.8 $\pm$ 0.4 & $<$0.76 & $<$0.60 \\
				$R_{2\rightarrow1}$ & 0.7 $\pm$ 0.1 & 0.5 $\pm$ 0.1 & $<$0.26 & $<$0.21 \\
				\hline
			\end{tabular}
			\caption{$z$ is the redshift, calculated as $z=\nu_{\mathrm{rest}}/\nu_{\mathrm{obs}} - 1$. $\nu_{\mathrm{obs}}$ refers to the observed frequency of the line emission peak. $S_{\mathrm{peak}}$ refers to the peak flux of the Gaussian fit. $S_{\mathrm{cont}}$ refers to the continuum flux. $I$ is the integrated flux and $L'$ is the line luminosity. $R_{2\rightarrow1}$ refers to the ratio of the line luminosities of the $5\rightarrow4$ and $6\rightarrow5$ lines to the $2\rightarrow1$ line. HZ10 data: for $S_{\mathrm{peak}}$, $S_{\mathrm{cont}}$ and $I$ we report 3 times the rms noise added to the measured value. The numbers for $L'$ and $R_{2\rightarrow1}$ are derived from these upper limits. Numbers in parentheses are adopted from CO(2$\rightarrow$1) measurements in Pavesi et al. (\hyperlink{Pa}{2019})}
			
			\centering
			\begin{tabular}{p{6cm} p{6cm}}
				\multicolumn{2}{c}{\textbf{Table 2}}\\
				\multicolumn{2}{c}{CRLE SED Parameters}\\
				\hline
				\hline
				$L_{\mathrm{IR}}$ ($10^{13} L_\odot$) & $3.0 \pm 0.1$\\
				$L_{\mathrm{FIR}}$ ($10^{13} L_\odot$) & $1.7 \pm 0.1$\\
				$T_d$ (K) & $57 \pm 4$\\
				$\lambda_0$ ($\mu$m) & $147^{+17}_{-22}$\\
				$\alpha$ & $2.1 \pm 0.1$\\
				$\beta$ & $2.1 \pm 0.1$\\
				\hline
			\end{tabular}
			\caption{SED parameters for CRLE: $L_{\mathrm{IR}}$ is the infrared luminosity, obtained by integrating between 8 and 1000 $\mu$m while $L_{\mathrm{FIR}}$ is the far-infrared luminosity, obtained by integrating between 42.5 and 122.5 $\mu$m. $T_d$ is the dust temperature, $\lambda_0$ is the rest-frame wavelength at which the optical depth becomes unity, $\alpha$ is a mid-IR power law index and $\beta$ is the dust emissivity parameter. All values were obtained by using the \texttt{mbb\_emcee} code (Riechers et al. \hyperlink{R2}{2013}, Dowell et al. \hyperlink{Do}{2014}) wrapped in an affine invariant MCMC package described in Foreman-Mackey et al. (\hyperlink{FM}{2013}).}
		\end{table}
	\end{turnpage}
	
	\clearpage
	
	\begin{table}
		\begin{tabular}{p{4cm} p{4cm} p{4cm} p{4cm}}
			\multicolumn{4}{c}{\textbf{Table 3}}\\
			\multicolumn{4}{c}{Measured Continuum Fluxes (CRLE)}\\
			\hline
			\hline
			$\lambda_{\rm{rest}}$ ($\mu$m) & $\nu_{\rm{obs}}$ (GHz) & Flux (mJy) & Band\\
			\hline
			37.5 & 1200 & 12.0 $\pm$ 0.9 & \textit{Herschel}/SPIRE\\
			52.5 & 857 & 20.9 $\pm$ 1.3 & ... \\
			75.0 & 600 & 31.1 $\pm$ 1.4 & ... \\
			127.5 & 353 & 16.3 $^{+1.6}_{-2.2}$ & SCUBA-2 \\
			153.6 & 293 & 16.5 $\pm$ 0.9 & ALMA \\
			196.2 & 229 & 8.65 $\pm$ 0.3 & ... \\
			433.6 & 104 & 0.534 $\pm$ 0.023 & NOEMA \\
			520.2 & 86.5 & 0.317 $\pm$ 0.025 & NOEMA \\
			1320 & 34.1 & 0.022 $\pm$ 0.002 & VLA \\
			\hline
		\end{tabular}
		\caption{Measured continuum fluxes for CRLE. Data for $\lambda_{\rm{rest}}$ = 37.5-196.2 $\mu$m are from P18 with the exception of the 127.5 $\mu$m SCUBA-2 value from Simpson et al. (\hyperlink{S19}{2019}; S2COSMOS ID COS850.33). The data points at 433.6 and 520.2 $\mu$m are measured by collapsing all the line-free channels in the sidebands containing the CO(6$\rightarrow$5) and CO(5$\rightarrow$4) transitions, respectively. The contribution of the foreground galaxy to the measured fluxes is expected to be negligible at these wavelengths.}
	\end{table}
	
	\clearpage
	
	\begin{figure}
		\centering
		\subfigure{\includegraphics[width=120mm]{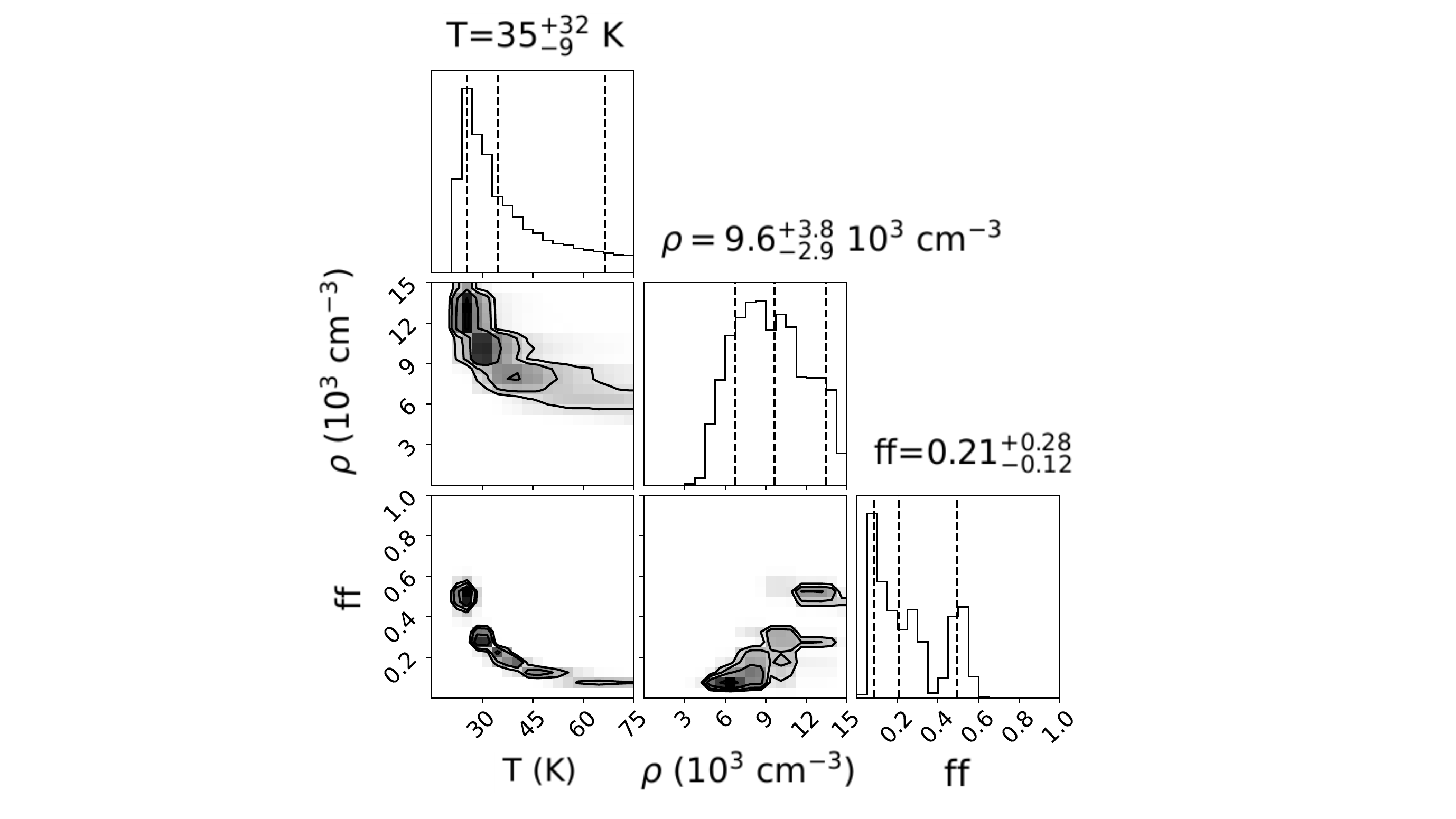}}
		\caption{A corner plot of the best-fit kinetic temperature, H$_2$ density and filling factor found by running an MCMC code on a grid of LVG results. The grid spans temperatures from 5-100 K and densities from 10$^{2}$-10$^{6.1}$ cm$^{-3}$. The filling factor is allowed to vary between 0 and 1. The physical size used as reference for the filling factor is the size of the [CII] emission reported in P18. Dashed lines correspond to the 16th, 50th and 84th percentiles.}
	\end{figure}
	
	\section{Appendix}
	
	We ran an MCMC code on a grid of LVG results to obtain the most likely values for the temperature, H$_2$ density and filling factor (see section 4 for details). Figure 6 shows the median value for each parameter, together with $\pm$1 $\sigma$ values and the overall distribution of parameter values found by the MCMC code.
	
	\clearpage
	
	\noindent
	\hypertarget{A}{Apostolovski, Y., Aravena, M., Anguita, T., et al. 2019, A\&A, 628, A23}
	\newline
	\hypertarget{Ar}{Aravena, M., Spilker, J. S., Bethermin, M., et al. 2016, MNRAS, 457, 4406}
	\newline
	\hypertarget{B}{Bethermin, M., Daddi, E., Magdis, G., et al. 2015, A\&A, 573, A113}
	\newline
	\hypertarget{Boo}{Boogaard, L. A., van der Werf, P., Weiss, A., et al. 2020, ApJ, 902, 109}
	\newline
	\hypertarget{Bo}{Bothwell, M. S., Smail, I., Chapman, S. C., et al. 2013, MNRAS, 429, 3047}
	\newline
	\hypertarget{C}{Capak, P. L., Carilli, C., Jones, G., et al. 2015, Natur, 522, 455}
	\newline
	\hypertarget{Cap}{Caputi, K. I., Dole, H., Lagache, G., et al. 2006, A\&A, 454, 143}
	\newline
	\hypertarget{C10}{Carilli, C. L., Daddi, E., Riechers, D., et al. 2010, ApJ, 714, 1407}
	\newline
	\hypertarget{CW}{Carilli, C. L., Walter, F., 2013, ARA\&A, 51, 105}
	\newline
	\hypertarget{Cas}{Casey, C. M., 2012, MNRAS, 425, 3094}
	\newline
	\hypertarget{Co}{Cortzen, I., Magdis, G. E., Valentino, F., et al. 2020, A\&A, 634, L14}
	\newline
	\hypertarget{dC}{da Cunha, E., Groves, B., Walter, F., et al. 2013, ApJ, 766, 13}
	\newline
	\hypertarget{Da}{Daddi, E., Dannerbauer, H., Liu, D., et al. 2015, A\&A, 577, A46}
	\newline
	\hypertarget{D}{Damjanov, I., McCarthy, P. J., Abraham, R. G., et al. 2009, ApJ, 695, 101}
	\newline
	\hypertarget{Do}{Dowell, C. D., Conley, A., Glenn, J., et al. 2014, ApJ, 780, 75}
	\newline
	\hypertarget{Dz}{Dudzeviciute, U., Smail, I., Swinbank, A. M., et al. 2020, MNRAS, 494, 3828}
	\newline
	\hypertarget{Fa}{Faisst, A. L., Fudamoto, Y., Oesch, P. A., et al. 2020, MNRAS, 498, 4192}
	\newline
	\hypertarget{FM}{Foreman-Mackey, D., Hogg, D. W., Lang, D., \& Goodman, J. 2013, PASP, 125, 306}
	\newline
	\hypertarget{F}{Fu, H., Cooray, A., Feruglio C., et al. 2013, Natur, 498, 338}
	\newline
	\hypertarget{GP}{Geach, J. E., Papadopoulos, P. P., 2012, ApJ, 757, 156}
	\newline
	\hypertarget{G}{Gruppioni, C., Bethermin, M., Loiacono, F., et al. 2020, A\&A, 643, A8}
	\newline
	\hypertarget{HG}{Hinojosa-Goni, R., Munoz-Tunon, C., Mendez-Abreu, J., 2016, A\&A, 592, A122}
	\newline
	\hypertarget{H12}{Hodge, J. A., Carilli, C. L., Walter, F., et al. 2012, ApJ, 760, 11}
	\newline
	\hypertarget{H}{Hodge, J. A., Swinbank, A. M., Simpson, J. M., et al. 2016, ApJ, 833, 103}
	\newline
	\hypertarget{J}{Jin, S., Daddi, E., Magdis, G. E., et al. 2019, ApJ, 887, 144}
	\newline
	\hypertarget{K}{Kennicutt, R. C., Jr. 1998, ApJ, 498, 541}
	\newline
	\hypertarget{N}{Narayanan, D., Krumholz, M. R., 2014, MNRAS, 442, 1411}
	\newline
	\hypertarget{Pav}{Pavesi, R., Riechers, D. A., Capak, P. L., et al. 2016, ApJ, 832, 151}
	\newline
	\hypertarget{Pa}{Pavesi, R., Riechers, D. A., Faisst, A. L., et al. 2019, ApJ, 882, 168}
	\newline
	\hypertarget{P}{Pavesi, R., Riechers, D. A., Sharon, C. E., et al. 2018, ApJ, 861, 43}
	\newline
	\hypertarget{Pop}{Popping, G., van Kampen, E., Decarli, R., et al. 2016, MNRAS, 461, 93}
	\newline
	\hypertarget{R2}{Riechers, D. A., Bradford, C. M., Clements, D. L., et al. 2013, Natur, 496, 329}
	\newline
	\hypertarget{R1}{Riechers, D. A., Capak, P. L., Carilli, C. L., et al. 2010, ApJL, 720, L131}
	\newline
	\hypertarget{R4}{Riechers, D. A., Hodge, J. A., Pavesi, R., et al. 2020, ApJ, 895, 81}
	\newline
	\hypertarget{R3}{Riechers, D. A., Leung, T. K. D., Ivison, R. J., et al. 2017, ApJ, 850, 1}
	\newline
	\hypertarget{R5}{Riechers, D. A., Pavesi, R., Sharon, C. E., et al. 2019, ApJ, 872, 7}
	\newline
	\hypertarget{Sc}{Schreiber, C., Elbaz, D., Pannella, M., et al. 2018, A\&A, 609, A30}
	\newline
	\hypertarget{S03}{Scoville, N., 2003, Journal of Korean Astronomical Society, 36, 167}
	\newline
	\hypertarget{Sco}{Scoville, N., Lee, N., Vanden Bout, P., et al. 2017, ApJ, 837, 150}
	\newline
	\hypertarget{Sco1}{Scoville, N., Sheth, K., Aussel, H., et al. 2016, ApJ, 820, 83}
	\newline
	\hypertarget{S19}{Simpson, J. M., Smail, I., Swinbank, A. M., et al. 2019, ApJ, 880, 43}
	\newline
	\hypertarget{S17}{Smol{\^c}i{\'c}, V., Novak, M., Bondi, M., et al. 2017, A\&A, 602, A1}
	\newline
	\hypertarget{Sp}{Spilker, J. S., Marrone, D. P., Aguirre, J. E., et al. 2014, ApJ, 785, 149}
	\newline
	\hypertarget{Sp2}{Spilker, J. S., Marrone, D. P., Aravena, M., et al. 2016, ApJ, 826, 112}
	\newline
	\hypertarget{S2}{Strandet, M. L., Weiss, A., De Breuck, C., et al. 2017, ApJL, 842, L15}
	\newline
	\hypertarget{S1}{Strandet, M. L., Weiss, A., Vieira, J. D., et al. 2016, ApJ, 822, 80}
	\newline
	\hypertarget{T05}{Thompson, T. A., Quataert, E., Murray, N., 2005, ApJ, 630, 167}
	\newline
	\hypertarget{V}{Vallini, L., Gruppioni, C., Pozzi, F., Vignali, C., \& Zamorani, G. 2016, MNRAS, 456, L40}
	\newline
	\hypertarget{Wa}{Walter, F., Decarli, R., Carilli, C., et al. 2012, Natur, 486, 233}
	\newline
	\hypertarget{W}{Wang, L., De Lucia, G., Fontanot, F., et al. 2018, MNRAS, 000, 1}
	\newline
	\hypertarget{We}{Weiss, A., De Breuck, C., Marrone, D. P., et al. 2013, ApJ, 767, 88}
	\newline
	\hypertarget{W07}{Weiss, A., Downes, D., Walter, F., et al. 2007, In \textit{ASP Conf. Ser. 375, From Z-Machines to ALMA: (Sub)Millimeter Spectroscopy of Galaxies}, p. 25}
	\newline
	\hypertarget{W05}{Weiss, A., Walter, F., Scoville, N. Z., 2005, A\&A, 438, 533}

\end{document}